\documentclass[12pt]{article}

\usepackage{graphicx}
\begin{document}

\newcommand{\siml}{\stackrel{<}{\sim}}
\newcommand{\simg}{\stackrel{>}{\sim}}
\newcommand{\lleq}{\stackrel{<}{=}}

\baselineskip=1.333\baselineskip


%
\begin{center}
{\large\bf
A moment approach to 
non-Gaussian colored noise
} 
\end{center}

\begin{center}
Hideo Hasegawa
\footnote{E-mail address:  hideohasegawa@goo.jp}
\end{center}

\begin{center}
{\it Department of Physics, Tokyo Gakugei University  \\
Koganei, Tokyo 184-8501, Japan}
\end{center}
\begin{center}
({\today})
\end{center}
\thispagestyle{myheadings}

\begin{abstract}

The Langevin system subjected to non-Gaussian noise
has been discussed, by using the second-order moment approach 
with two kinds of models for generating the noise.
We have derived the effective differential equation (DE) for
a variable $x$, from which the stationary probability distribution $P(x)$
has been calculated with the use of the Fokker-Planck equation.
The result of $P(x)$ calculated by the moment method 
is compared to several expressions obtained by
different methods such as the universal
colored noise approximation (UCNA) 
[Jung and H\"{a}nggi, Phys. Rev. A {\bf 35}, 4464 (1987)]
and the functional-integral method.
It has been shown that our $P(x)$ 
is in good agreement with that of direct simulations (DSs).
We have also discussed
dynamical properties of the model with an external input,
solving DEs in the moment method.

\end{abstract}

\vspace{0.5cm}

\vspace{3.0cm}

{\rm Phone: +81-42-329-7482}

{\rm Fax: +81-42-329-7491}

{\rm E-mail: hideohasegawa@goo.jp}

\newpage
\section{INTRODUCTION}

Interesting, unconventional phenomena 
such as the stochastic resonance (SR) and the noise-induced
phase transition are created by noise. Theoretical studies on noise 
in nonlinear dynamical systems
have usually adopted Gaussian white (or colored) noise.
In recent years, there is a growing interest in studying dynamical 
systems driven by non-Gaussian noise.
This is motivated by the fact that non-Gaussian noise
with random amplitudes following the power-law distribution
is quite ubiquitous in natural phenomena. 
For example, experimental results for crayfish and rat skin
offer strong indication that there could be
non-Gaussian noise in these sensory systems 
\cite{Wiesen94}\cite{Nozaki99}.
A simple mechanism has been proposed
to generate the non-Gaussian noise \cite{Borland98}.
With the use of such a theoretical model,
the SR induced by non-Gaussian colored noise
has been investigated \cite{Fuentes01}. 
It has been shown that the peak in the signal-to-noise ratio (SNR)
for non-Gaussian noise becomes broader than that for Gaussian noise.
This result has been confirmed by an analog experiment \cite{Castro01}.

Stochastic systems with non-Gaussian colored noise 
are originally expressed by the non-Markovian process.
This problem is transformed into a Markovian one
by extending the number of variables and equations.
The relevant Fokker-Planck equation (FPE) includes 
the probability distribution expressed in terms of multi-variables.
We may transform this FPE for multivariate probability to
the effective single-variable FPE, or obtain 
one-variable differential equation (DE)
with the use of some approximation methods like 
the universal colored noise approximation (UCNA) 
\cite{Jung87,Hanggi95}
and the functional-integral
methods \cite{Fuentes02}\cite{Wu07}.  
The obtained results, however, do not agree each other,
depending on the adopted approximations, 
as will be explained in Sec. 2.2 (Table 1). 
It is not easy to
trace the origin of this discrepancy because of the
complexity in adopted procedures.
The purpose of 
the present paper is to discuss the non-Gaussian noise
and to make a comparison among various methods,
by employing the second-order moment method
which is simple and transparent, and which is exact
in the weak-noise limit.

The paper is organized as follows.
We have applied the second-moment method to the Langevin model
subjected to non-Gaussian noise which is generated
by two kinds of models.
In Sec. 2, non-Gaussian noise is generated 
by the specific function which was proposed by Borland \cite{Borland98}
and which has been adopted in several studies 
\cite{Fuentes01}\cite{Fuentes02}\cite{Wu07}.
In contrast, in Sec. 3,
non-Gaussian noise is generated by multiplicative noise
\cite{Sakaguchi01}-\cite{Hasegawa07}.
We derive the effective one-variable DE, from which
the stationary distribution is calculated with the use of
the FPE.
A comparison among various methods generating the 
non-Gaussian noise is made is Sec. 4, where contributions
from higher moments than the second moment
are also discussed.
The final Sec. 5 is devoted to our conclusion.

\section{Models A$_0$ and A}

\subsection{Moment method}

We have adopted the Langevin model subjected to
non-Gaussian colored noise ($\epsilon$) 
and Gaussian white noise ($\psi \xi$), 
as given by \cite{Borland98}
\begin{eqnarray}
\dot{x}&=& F(x) + \epsilon(t) + \psi \xi(t)+I(t), \\
\tau \dot{\epsilon} & =& K(\epsilon)
+ \phi \eta(t),
\hspace{2cm}\mbox{(model A$_0$)}
\end{eqnarray}
with
\begin{equation}
K(\epsilon)=- \frac{\epsilon}{[1+(q-1)(\tau/\phi^2) \epsilon^2]},
\end{equation}
which is referred to as the model A$_0$.
In Eqs. (1)-(3), $F(x)$ is an arbitrary function of $x$,
$I(t)$ stands for an external input,
$q$ is a parameter expressing a departure from
the Gaussian distribution which is realized for $q=1$,
$\tau$ denotes the characteristic time of colored noise,
and $\eta$ and $\xi$ the zero-mean white noises
with correlations: 
$\langle \eta(t) \eta(t') \rangle=\delta(t-t')$,
$\langle \xi(t) \xi(t') \rangle=\delta(t-t')$ and
$\langle \eta(t) \xi(t') \rangle=0$.

First, we briefly discuss the non-Gaussian colored noise
generated by Eqs. (2) and (3), 
which yield 
the stationary distribution given by \cite{Borland98}
\cite{Tsallis88,Tsallis98}
\begin{equation}
p_q(\epsilon) \propto \left[1+(q-1)
\left(\frac{\tau}{\phi^2} \right)\epsilon^2\right]_{+}^{-\frac{1}{q-1}},
\end{equation}
with $[x]_{+}=x$ for $x \geq 0$ and zero otherwise.
For $q=1$, Eq. (3) reduces to
\begin{equation}
K(\epsilon)=- \epsilon,
\end{equation}
which leads to the Gaussian distribution given by
\begin{equation}
p_1(\epsilon) \propto e^{-(\tau/\phi^2)\epsilon^2}.
\end{equation}
For $q > 1$ and $q < 1$, Eq. (4) yields long-tail and
cut-off distributions, respectively.
Thus Eqs. (2) and (3) generate the Gaussian and non-Gaussian noises,
depending on the value of parameter $q$.
Expectation values of $\epsilon$ and $\epsilon^2$ are given by
\begin{eqnarray}
\langle \epsilon \rangle &=& 0,\\
\langle \epsilon^2 \rangle 
&=& \frac{\phi^2}{\tau (5-3 q)},
\end{eqnarray}
which shows that $\langle \epsilon^2 \rangle$ 
diverges at $q=5/3$.

In order to make our calculation tractable, we 
replace the $\epsilon^2$ term in the denominator of Eq. (3) 
by its expectation value: 
$\epsilon^2 \simeq \langle \epsilon^2 \rangle$, 
to get \cite{Wu07}
\begin{eqnarray}
K(\epsilon) &\simeq& - \frac{\epsilon}{r_q},
\end{eqnarray}
by which Eq. (2) becomes
\begin{eqnarray}
\tau \dot{\epsilon} & =& - \left( \frac{1}{r_q} \right)\:\epsilon 
+ \phi \:\:\eta(t),
\hspace{2cm}\mbox{(model A)}
\end{eqnarray}
with
\begin{eqnarray}
r_q &=& \frac{2(2-q)}{(5-3q)}.
\end{eqnarray}
A model given by Eq. (1) with Eq. (10) is hereafter 
referred to as the model A, which is discriminated
from the model A$_0$ given by Eqs. (1)-(3).
The solid curve in Fig. 1 expresses $r_q$.
We note that we get $r_q=1$
for $q=1$, and 
$r_q < 1$ ($r_q > 1$) for $q < 1$ ($1 < q < 5/3$).
The dashed curve will be discussed in Sec. 3.1.

Now we discuss 
the FPE of the distribution $p(x, \epsilon, t)$ for
Eqs. (1) and (10), which are regarded as
the coupled Langevin model.
We get
\begin{eqnarray}
\frac{\partial}{\partial t} p(x,\epsilon,t) &=& 
-\frac{\partial }{\partial x}
\{[F(x)+ \epsilon + I] p(x,\epsilon,t)\} 
+ \frac{\psi^2}{2} \frac{\partial^2}{\partial x^2}
p(x,\epsilon,t) \nonumber \\
&+&\frac{1}{r_q \tau}\frac{\partial}{\partial \epsilon} 
[\epsilon p(x,\epsilon,t)] 
+ \frac{1}{2}\left(\frac{\phi}{\tau} \right)^2 
\frac{\partial}{\partial \epsilon} \left[\epsilon
\frac{\partial}{\partial \epsilon} \epsilon p(x,\epsilon,t) \right].
\end{eqnarray}
We define means, variances and covariances by 
\begin{eqnarray}
\langle x^m \epsilon^n \rangle 
&=& \int dx \int d \epsilon \; x^m \epsilon^n p(x,\epsilon,t).
\hspace{1cm}\mbox{($m, n$: integer)}
\end{eqnarray}
By using the moment method for the coupled Langevin model
\cite{Hasegawa06,Hasegawa07},
we get their equations of motion given by 
\begin{eqnarray}
\frac{d \langle x \rangle}{d t} 
&=& \langle F(x) + \epsilon + I \rangle, \\
\frac{d \langle \epsilon \rangle}{d t} 
&=& - \frac{1}{r_q \tau}\langle  \epsilon \rangle, \\
\frac{d \langle x^2 \rangle}{d t} 
&=& 2 \langle x[F(x) + \epsilon + I] \rangle + \psi^2, \\
\frac{d \langle \epsilon^2 \rangle}{d t} 
&=& - \frac{2}{r_q \tau}\langle \epsilon^2 \rangle
+\left( \frac{\phi}{\tau} \right)^2, \\
\frac{d \langle x \epsilon \rangle}{dt}
&=& \langle \epsilon [F(x) + \epsilon + I] \rangle
-\frac{1}{r_q \tau}\langle x \epsilon \rangle.
\end{eqnarray}

We consider means, variances and covariance defined by
\begin{eqnarray}
\mu &=& \langle x \rangle, \\
\nu &=& \langle \epsilon \rangle, \\
\gamma &=& \langle x^2 \rangle - \langle x^2 \rangle^2, \\
\zeta &=& \langle \epsilon^2 \rangle - \langle \epsilon \rangle^2, \\
\chi &=& \langle x \epsilon \rangle
- \langle x \rangle \langle \epsilon \rangle.
\end{eqnarray}
When we expand Eqs. (14)-(18) 
as $x=\mu+\delta x$ and $\epsilon=\nu+\delta \epsilon$
around the mean values of $\mu$ and $\nu$,
and retaining up to their second order 
contributions such as $ \langle (\delta x)^2 \rangle $, 
equations of motion become \cite{Hasegawa06,Hasegawa07}
\begin{eqnarray}
\frac{d \mu}{dt} &=& f_0 + f_2 \gamma
+ \nu  +I(t), \\
\frac{d \nu}{dt} &=& - \frac{\nu}{r_q \tau}, \\
\frac{d \gamma}{dt} &=& 2(f_1 \gamma + \phi)
+ \psi^2, \\
\frac{d \zeta}{dt} &=& - \left( \frac{2}{r_q \tau} \right) \zeta
+ \left( \frac{\phi}{\tau} \right)^2, \\
\frac{d \chi}{dt} &=& \left(f_1 -\frac{1}{r_q \tau} \right) \chi+ \zeta,
\end{eqnarray}
with
\begin{eqnarray}
f_{\ell} &=& \frac{1}{\ell !} 
\frac{\partial^{\ell} F(\mu)}{\partial x^{\ell}}. 
\end{eqnarray}

When we adopt the stationary values
for $\nu$, $\zeta$ and $\phi$:
\begin{eqnarray}
\nu& \simeq &\nu_s  =0, \\
\zeta & \simeq&\zeta_s =\frac{r_q \phi^2}{2 \tau}, \\
\chi &\ \simeq&\chi_s = \frac{r_q^2 \phi^2}{2(1-r_q\tau f_1)},
\end{eqnarray}
equations of motion for 
$\mu$ and $\gamma$ become
\begin{eqnarray}
\frac{d \mu}{dt} &=& f_0 + f_2 \gamma + I(t), \\
\frac{d \gamma}{dt} &=& 2 f_1 \gamma 
+ \frac{r_q^2 \phi^2}{(1-r_q \tau f_1)}+\psi^2, 
\end{eqnarray}
where $r_q$ is given by Eq. (11).
It is noted that the stationary value of 
$\zeta_s=(2-q)\phi^2/\tau(5-3q)$ in Eq. (31) is a little different
from $\langle \epsilon^2 \rangle=\phi^2/\tau(5-3q)$ in Eq. (8), 
which is due to an introduced approximation.

We may express the effective DE for $x$ as
\begin{eqnarray}
\dot{x}&=& F_{eff}(x) + I_{eff}(t) + \alpha_{eff} \:\eta(t) 
+ \psi \xi(t),
\end{eqnarray}
with
\begin{eqnarray}
F_{eff}(x) &=& F(x), \\
I_{eff}(t) &=& I(t), \\
\alpha_{eff} &=& \frac{\phi_q}{\sqrt{1- \tau_q f_1}}, \\
\phi_q &=& r_q \phi, \\
\tau_q &=& r_q \tau,
\end{eqnarray}
from which Eqs. (33) and (34) are derived \cite{Hasegawa06,Hasegawa07}.
Equations (35) and (38) clearly express the effect of non-Gaussian
colored noise. The effective magnitude of noise $\alpha_{eff}$
is increased with increasing $q$ (Fig. 1). In contrast,
with increasing $\tau$, the effective $\alpha_{eff}$ value is
decreased for $f_1 < 0$ which is usually realized.

The FPE of $P(x,t)$ for Eq. (35) is expressed by
\begin{eqnarray}
\frac{\partial }{\partial t}P(x,t) &=& 
- \frac{\partial}{\partial \epsilon} 
\{ (F_{eff}+I) P(x,t) \}
+ \frac{1}{2}
\frac{\partial}{\partial x} \left[ \alpha_{eff} \frac{\partial}
{\partial x} \alpha_{eff} P(x,t) \right] \nonumber \\
&+& \frac{\psi^2}{2}
\frac{\partial^2}{\partial x^2}P(x,t),
\end{eqnarray}
which may be applicable to $\alpha_{eff}$ depending on $x$
[{\it i.e.} Eqs. (50) and (59)].
The stationary distribution is given by
\begin{eqnarray}
\ln P(x) &=& 2 \int \:dx
\left( \frac{F_{eff}+I}{\alpha_{eff}^2+\psi^2} \right) 
-\frac{1}{2} \ln \left(\frac{\alpha_{eff}^2+\psi^2}{2} \right).
\end{eqnarray}
For $F(x)=-\lambda x$, we get
\begin{eqnarray}
P(x) &\propto & \exp\left[-\left(\frac{\lambda}
{\left[\phi_q^2/(1+\lambda \tau_q)+\psi^2 \right]}\right)
\left( x-\frac{I}{\lambda} \right)^2\right].
\end{eqnarray}
For $F(x)=a x - b x^3$, we get
\begin{eqnarray}
P(x) &\propto & \exp\left[\frac{1}
{[\phi_q^2/(1-\tau_q(a - 3 b\mu^2))+\psi^2]}\right]
\left(ax^2-\frac{bx^4}{2}+2 Ix \right).
\end{eqnarray}

\subsection{Comparison with other methods}
We will compare the result of the moment method with those of several 
analytical methods: the universal colored noise approximation
(UCNA) and functional-integral methods (FI-1 and FI-2).

\noindent
{\bf (a) UCNA}

The universal colored noise approximation (UCNA)
was proposed by
Jung and H\"{a}nggi \cite{Jung87,Hanggi95}
by interpolating between
the two limits of $\tau=0$ and $\tau=\infty$ of colored noise,
and it has been widely adopted for a study
of effects of Gaussian and non-Gaussian colored noises.
By employing the UCNA, 
we may derive the effective DE for the variable $x$.
Taking the time derivative of Eq. (1) with $\psi=0$, and
using Eq. (10) for $\dot{\epsilon}$, we get
\begin{eqnarray}
\ddot{x} &=& F' \dot{x}+ \dot{\epsilon} + \dot{I}, \\
&=& \left(F'-\frac{1}{\tau_q}\right) \dot{x}
+ \left(\frac{F+I}{\tau_q} \right)
+\dot{I}+\left( \frac{r_q \phi}{\tau_q}\right) \eta.
\end{eqnarray}
When we neglect the $\ddot{x}$ term after the UCNA,
we get the effective DE for $x$ given by
\begin{eqnarray}
\dot{x}&=& F_{eff}(x)+ I_{eff}(t) + \alpha_{eff} \:\eta(t),
\end{eqnarray}
with
\begin{eqnarray}
F^U_{eff}(x) &=& \frac{F(x)}{(1 - \tau_q F')}, \\
I^U_{eff}(t)  &=& \frac{(I+\tau_q \dot{I})}{(1 - \tau_q F')}, \\
\alpha^U_{eff} &=& \frac{r_q \phi}{(1 - \tau_qF')},
\end{eqnarray}
where $F=F(x)$, $F'=F'(x)$, and
$\tau_q$ and $r_q$ are given by Eqs. (40) and (11), respectively.
It is noted that $\alpha^U_{eff}$ given by Eq. (50)
generally depends on $x$, yielding the multiplicative 
noise in Eq. (47).

For $F(x)=-\lambda x$, the stationary distribution is given by
\begin{eqnarray}
P^U(x) &\propto & \exp\left[-\frac{\lambda (1+\lambda \tau_q)}{\phi_q^2}
\left(x-\frac{I_c}{\lambda} \right)^2 \right],
\end{eqnarray}
which agrees with the result of Eq. (43) with $\psi=0$.

For $F(x)=ax-b x^3$, we get 
\begin{eqnarray}
P^U(x) &\propto & [1-\tau_q(a-3bx^2)]
\exp \left( \left[\frac{a(1-a\tau_q)x^2}{\phi_q^2} 
-\frac{b(1-4a\tau_q)x^4}{2\phi_q^2}
-\frac{b^2\tau_q x^6}{\phi_q^2} \right] \right)
\nonumber \\
&&\times \exp\left( \left[\frac{2I_c(1-a\tau_q)x}{\phi_q^2}
+\frac{2I_cb\tau_q x^2}{\phi_q^2} \right] \right),
\end{eqnarray}
whose functional form is rather different from that given by Eq. (44).

\noindent
{\bf (b) Functional-integral method (FI-1)}

Wu, Luo and Zhu \cite{Wu07} started from the formally 
exact expression for $P(x,t)$
of Eqs. (1) and (10) with $I(t)=0$ given by
\begin{eqnarray}
\frac{\partial}{\partial t} P(x,t)
&=& -\frac{\partial}{\partial x} [F(x) P(x,t)]
-\frac{\partial}{\partial x}  
\langle \epsilon(t) \delta(x(t)-x) \rangle
-\psi \frac{\partial}{\partial x}  
\langle \xi(t) \delta(x(t)-x) \rangle,
\end{eqnarray}
where $\langle \cdot \rangle $ denotes the average
over the probability $P(x,t)$ to be determined.
Employing the Novikov theorem \cite{Novikov65} and
the functional-integral method, they obtained
the effective FPE for $P(x,t)$ which yields Eq. (47) but with
\begin{eqnarray}
F^W_{eff}(x) &=& F(x), \\
\alpha^W_{eff} &=& \frac{r_q \phi}
{\sqrt{1- \tau_q F_s'} },
\end{eqnarray}
where $r_q$ and $\tau_q$ are given by Eqs. (40) and (11),
respectively,
$F'=dF/dx$ and
$F_s$ {\it et al.} denote the steady-state values at $x=x_s$.

For $F(x)=-\lambda x$, we get
\begin{eqnarray}
P^W(x) &\propto & \exp\left[-\frac{\lambda (1+\lambda \tau_q)}{\phi_q^2}
x^2 \right].
\end{eqnarray}
For $F(x)=ax-b x^3$, we get 
\begin{eqnarray}
P^W(x) &\propto&
\exp\left[\frac{[1-\tau_q(a-3b\mu^2)]}{\phi_q^2}
\left(ax^2-\frac{bx^4}{2} \right) \right].
\end{eqnarray}

\noindent
{\bf (c) Functional-integral method (FI-2)}

By applying the alternative functional-integral method
to the FPE for $p(x,\epsilon,t)$ given by Eqs. (1) and (10) 
with $\psi=I(t)=0$, 
Fuentes, Toral and Wio \cite{Fuentes01}
derived the FPE of $P(x,t)$, which leads to
the effective DE given by Eq. (47), but with
\begin{eqnarray}
F^F_{eff}(x) &=& \frac{F}{(1-s_q \tau  F') }, \\
\alpha^F_{eff} &=& \frac{ s_q \phi}
{(1- s_q \tau  F')}, 
\end{eqnarray}
with
\begin{equation}
s_q= \left[1+(q-1)\left(\frac{\tau}{2\phi^2}\right)F^2\right].
\end{equation}
We note that $\alpha^F_{eff}$
generally depends on $x$, yielding the multiplicative noise in Eq. (47).

For $F(x)=-\lambda x$, we get
\begin{eqnarray}
P^F(x) &\propto & 
(1+\lambda \tau s_q ) s_q^{-[2/(q-1)+1]}
\exp\left[ \frac{2}{\lambda \tau (q-1) s_q} \right],
\end{eqnarray}
with
\begin{equation}
s_q= \left[1+(q-1)\left(\frac{\tau \lambda^2}{2\phi^2}\right)x^2 \right].
\end{equation}
For $F(x)=a x - b x^3$, it is necessary to numerically
evaluate the distribution $P(x)$ 
with the use of Eqs. (42) and (58)-(60).

A comparison among various methods is summarized in the Table 1.
We note that the result of our moment method agrees with that of
FI-1, but disagrees with those of UCNA and FI-2.
The result of UCNA is not identical with that
of FI-2, although they are consistent each other
if the identity of $s_q=r_q$ holds, 
which is realized for $q=1$ with $s_1=r_1=1$.

\subsection{Numerical calculations}

We present some numerical examples to make a comparison
with direct simulation (DS), which has been performed
for Eqs. (1)-(3) by the fourth-order Runge-Kutta method
with a time step of 0.01 for 1000 trials.
Figures 2(a)-2(f) show the stationary probability
calculated by various methods for $F(x)=-\lambda x$
with changing
parameters of $q$ and $\tau$ for fixed $\phi=0.5$
and $\psi=0$.
A comparison between Fig. 2(c) and 2(d) shows that
the width of the distribution for $\tau=1.0$ is narrower
than that for $\tau=0.5$.
This is explained by the reduced effective strength of
$\alpha_{eff}=\phi/(1+\lambda \tau)$ by an increased $\tau$.
We note that for $q=1.0$,
results of all methods are in good agreement each other.
Comparing Fig. 2(a) to Fig. 2(c) [and Fig. 2(b) to Fig. 2(d)],  
we note that the width of the distribution for $q=0.8$ is
a little narrower than that for $q=1.0$.
This is due to the fact that the $r_q$ value is reduced 
to 0.82 from unity.
An agreement among various methods is good for $q=0.8$.
In contrast, Figs. 2(e) and 2(f) show that for $q=1.5$,
the width of $P(x)$ becomes wider because of the increased 
$r_{1.5}=2.0$.
The results of the moment method, UCNA and FI-1
are in fairly good agreement.  
On the contrary, the distribution calculated by FI-2 is 
sharper than that of DS.

Figures 3(a)-3(f) show the stationary probability
calculated by various methods for $F(x)=x-x^3$
with changing
parameters of $q$ and $\tau$ for fixed $\phi=0.5$
and $\psi=0.0$.
The general trend realized in Figs. 3(a)-3(f) is the same
as in Figs.2(a)-2(f).
The result of FI-2 for $q=1.5$ is not so bad compared to
those of other approximation methods. However, the result
of FI-2 for $q=0.8$ and $\tau=1.0$ is worse than
other methods.

\section{Model B}

\subsection{Moment method}

In order to generate non-Gaussian noise,
we may employ an alternative model
(referred to as the model B) given by
\begin{eqnarray}
\dot{x}&=& F(x) + \epsilon(t) +I(t), \\
\tau \dot{\epsilon} & =& -\epsilon 
+ \epsilon \:\alpha \eta(t) + \beta \xi(t),
\hspace{2cm}\mbox{(model B)}
\end{eqnarray}
where $F(x)$ expresses an arbitrary function of $x$,
$I(t)$ an external input,
$\tau$ the characteristic time of colored noise, and
$\alpha$ and $\beta$ denote magnitudes of
additive and multiplicative noises, respectively, given by
zero-mean white noises, $\eta$ and $\xi$, with
correlations: 
$\langle \eta(t) \eta(t') \rangle
= \langle \xi(t) \xi(t') \rangle= \delta(t-t')$ and
$\langle \eta(t) \xi(t') \rangle=0$.

The FPE for the distribution $p(\epsilon, t)$ for Eq. (64)
in the Stratonovich representation is given by
\begin{eqnarray}
\frac{\partial}{\partial t} p(\epsilon,t) &=& 
\frac{1}{\tau}\frac{\partial}{\partial \epsilon} 
[\epsilon p(\epsilon,t)]
+ \frac{1}{2}\left(\frac{\alpha}{\tau}\right)^2 
\frac{\partial}{\partial \epsilon}\left( \epsilon
\frac{\partial}{\partial \epsilon} [\epsilon p(\epsilon,t)] \right)
+ \frac{1}{2}\left(\frac{\beta}{\tau}\right)^2 
\frac{\partial^2}{\partial \epsilon^2}
p(\epsilon,t).
\end{eqnarray}
The stationary distribution of $\epsilon$
has been extensively discussed 
\cite{Sakaguchi01}-\cite{Hasegawa07}
in the context of the
nonextensive statistics \cite{Tsallis88,Tsallis98}.
It is given by 
\cite{Sakaguchi01}-\cite{Hasegawa07}
\begin{eqnarray}
p_q(\epsilon)
&\propto& \left[1 +  \left(\frac{\alpha^2}{\beta^2}\right) 
\epsilon^2 \right]_{+}^{-(\tau/\alpha^2+1/2)}, \\
&\propto&
\left[1 + (q-1) 
\left(\frac{\tau}{\kappa \beta^2} \right)
\epsilon^2\right]_{+}^{-\frac{1}{q-1}},
\end{eqnarray}
with
\begin{eqnarray}
q&=& 1+ \left( \frac{2 \alpha^2}{2 \tau+\alpha^2} \right), \\
\kappa &=& \left( \frac{3-q}{2} \right)
= \left( \frac{2 \tau}{2 \tau+\alpha^2} \right),
\end{eqnarray}
where $[x]_{+}=x$ for $x \geq 0$ and zero otherwise.
In the limit of $\alpha=0.0$ ($q=1$), the distribution given by Eq. (67)
reduces to the Gaussian distribution given by
\begin{eqnarray}
p(\epsilon) &\propto& \exp\left(-\frac{\tau}{\beta^2} \epsilon^2\right).
\end{eqnarray}
In the opposite limit of $\beta=0.0$, Eq. (67) leads to the power-law
distribution given by
\begin{eqnarray}
p(\epsilon) &\propto& \epsilon^{- \delta}, 
\end{eqnarray}
with
\begin{eqnarray}
\delta &=&1+ \frac{2 \tau }{\alpha^2} = \frac{2}{q-1}. 
\end{eqnarray}
The expectation values of $\epsilon$ and $\epsilon^2$ are given by
\begin{eqnarray}
\langle \epsilon \rangle &=& 0, \\
\langle \epsilon^2 \rangle
&=& \frac{\kappa \beta^2}{\tau (5-3q)}
= \frac{\beta^2}{2(\tau-\alpha^2)}.
\end{eqnarray}
The second moment is finite for $\alpha^2 < \lambda$ ($q < 5/3$).
It is expected that
Eq. (64) leads to the non-Gaussian colored noise
with the correlation given by
\begin{equation}
\langle \epsilon(t) \epsilon(t') \rangle
= \frac{\beta^2}{2 (\tau-\alpha^2)} 
\exp\left[-\frac{\mid t-t' \mid}{\tau} \right].
\end{equation}

By applying the moment method to Eqs. (63) and (64),
we may obtain the effective one-variable DE for $x$
given by 
\begin{eqnarray}
\dot{x} &=& F_{eff}+ I(t)+ \beta_{eff}\; \xi(t),
\end{eqnarray}
with
\begin{eqnarray}
F_{eff} &=& F(x), \\
\beta_{eff} &=&  \frac{\beta_q}{\sqrt{1-\tau f_1}}, \\
\beta_q &=& \beta u_q, \\
u_q &=& \sqrt{\frac{1}{(1-\alpha^2/\tau)}}
= \sqrt{ \frac{3-q}{5-3q}},
\end{eqnarray}
details of calculations being explained in the Appendix.
The $q$ dependence of $u_q$ is plotted by 
the dashed curve in Fig. 1, where 
$u_q < 1$, $u_q=1$ and $u_q > 1$
for  $q <1$, $q=1$ and $1 < q < 5/3$, respectively.
We note that $u_q$ has a similar $q$ dependence 
as $r_q$ shown by the solid curve.


The FPE of $P(x,t)$ for Eq. (76) is given by
\begin{eqnarray}
\frac{\partial }{\partial t}P(x,t) &=& 
- \frac{\partial}{\partial \epsilon} 
\left[ (F_{eff}+I) P(x,t) \right]
+ \left(\frac{1}{2} \right) 
\frac{\partial}{\partial x}\left[\beta_{eff} \frac{\partial}
{\partial x}\beta_{eff} P(x,t)\right].
\end{eqnarray}
The stationary distribution is given by
\begin{eqnarray}
\ln P(x) &=& 2 \int \:dx
\left( \frac{F_{eff}+I}{\beta_{eff}^2} \right) 
-\frac{1}{2} \ln \left(\frac{\beta_{eff}^2}{2} \right).
\end{eqnarray}
For $F(x)=-\lambda x$, we get
\begin{eqnarray}
P(x) &\propto & \exp\left[-\left(\frac{\lambda \left(1+\lambda \tau \right)}
{\beta_q^2}\right)
\left( x-\frac{I}{\lambda} \right)^2\right].
\end{eqnarray}
For $F(x)=a x - b x^3$, we get
\begin{eqnarray}
P(x) &\propto & \exp\left[\frac{[1-\tau (a - 3 b\mu^2)]}
{\beta_q^2}\right]
\left(a x^2-\frac{bx^4}{2}+2 Ix \right).
\end{eqnarray}
Equations (83) and (84) are similar to Eqs. (43) and (44) 
(with $\psi=0.0$), respectively, for the model A, 
although $u_q$ and $\tau$ in the former are
different from $r_q$ and $\tau_q$ in the latter.

It would be interesting to compare the result of the moment
method for the model B with those of the UCNA and FI method,
as we have made for the model A in Sec. 2.2.
Unfortunately the UCNA method cannot be applied to
the model B because Eq. (64) 
includes the multiplicative noise \cite{Note1}.
It is very difficult to apply the FI method to the model B
including both additive and multiplicative noises:
such calculations have not been reported as far as
we are aware of.
Then we will make a comparison of the result
of the moment method
only with that of DS in the next subsection 3.2.

\subsection{Numerical calculations}

We present some numerical examples to make a comparison
with DS, which has been performed
for Eqs. (63) and (64) by the Heun method
with a time step of 0.001 for 1000 trials. 
Figures 4(a)-4(d) show the stationary probability $P(x)$
calculated for $F(x)=-\lambda x$ with changing
parameters of $\alpha$ and $\tau$ for a fixed $\beta=0.5$.
In Figs. 4(a) and 4(b) for $\alpha=0.0$ ($q=1.0$),
we observe that the width of $P(x)$ is decreased with increasing $\tau$,
as shown in Figs. 2(c) and 2(d). Figures 4(c) and 4(d) show that
with increasing $\alpha$ to 0.5, we get wider width in $P(x)$ because
we get $q=1.40$ and 1.22 for $\tau=0.5$ and 1.0, respectively [Eq. (68)].

Similarly, Figs. 5(a)-5(d) show $P(x)$ for $F(x)=x-x^3$.
Results of the moment method are in good agreement with those
of DS for $\alpha=0.0$ ($q=1.0$) as shown in Figs. 5(a) and 5(b). 
The width of $P(x)$ in Fig. 5(c) 
for $\alpha=0.5$ and $\tau=0.5$ ($q=1.40$)
is wider than that in Fig. 5(a) 
for $\alpha=0.0$ and $\tau=0.5$ ($q=1.0$),
but it is narrower than that in Fig. 5(d) 
for $\alpha=0.5$ and $\tau=1.0$ ($q=1.22$).

\section{Discussion}

We will make a comparison among the various methods
for generating non-Gaussian noise given by
\begin{eqnarray}
\dot{x}&=& F(x) + \epsilon(t) +I(t), 
\end{eqnarray}
with
\begin{eqnarray}
\tau \dot{\epsilon} & =& K(\epsilon)
+ \phi \eta(t),
\hspace{2cm}\mbox{(model A$_0$)} \\
\tau \dot{\epsilon} &=& -\left( \frac{\epsilon}{r_q} \right)
+\phi \eta(t),
\hspace{2cm}\mbox{(model A)} \\
\tau \dot{\epsilon} & =& -\epsilon 
+ \epsilon \:\alpha \eta(t) + \beta \xi(t),
\hspace{1cm}\mbox{(model B)}
\end{eqnarray}
where $\eta$ and $\xi$ are white noises, $r_q$ is given by Eq. (11),
and $K(\epsilon)$ is given by Eq. (3) or
\begin{eqnarray}
K(\epsilon) &=& - \frac{d U(\epsilon)}{d \epsilon}, \\
U(\epsilon) &=& \frac{\phi^2}{2 \tau (q-1) }
\ln \left[1 + (q-1) \left( \frac{\tau}{\phi^2} \right) 
\epsilon^2 \right].
\end{eqnarray}
Note that the model A is derived from the model A$_0$ 
with the approximation:
$K(\epsilon) \simeq -\epsilon/r_q$
and $U(\epsilon) \simeq \epsilon^2/2 r_q$ [Eq. (9)].
Noises in the models A$_0$ and A are generated by a motion
under the potentials given by Eq. (90)
and $U(\epsilon) = \epsilon^2/2 r_q$,
respectively, subjected to additive noise.
In contrast, noise in the model B is generated 
by a motion under the potential of $U(\epsilon)=\epsilon^2/2$
subjected to additive and multiplicative noises.

We note from Eqs. (4) and (67) that
the stationary distributions of $\epsilon$
in the models A$_0$ and B become the equivalent non-Gaussian
distribution if the
parameters in the two models satisfy the relation:
\begin{eqnarray}
\phi^2 &=& \kappa \beta^2 
= \frac{\beta^2}{(1+ \alpha^2/2 \tau)}.
\hspace{1cm}\mbox{for $q \geq 1$}
\end{eqnarray}
This equivalence, however, does no hold
between the models A and B, because
the stationary distribution of the model A
is not the non-Gaussian but the Gaussian given by
\begin{eqnarray}
p_q(\epsilon) \propto 
\exp \left[-\left(\frac{\tau}{r_q \phi^2} \right) \epsilon^2 \right].
\hspace{1cm}\mbox{(model A)} 
\end{eqnarray}

As for the dynamical properties,
equations of motion for $\langle \epsilon^2 \rangle$
in the moment method are given by
Eqs. (17) and (A6):
\begin{eqnarray}
\frac{d \langle \epsilon^2 \rangle}{dt} 
&=& - \left( \frac{2}{r_q \tau} \right) \langle \epsilon^2 \rangle
+ \left( \frac{\phi}{\tau} \right)^2, 
\hspace{1cm}\mbox{(model A)} \\
\frac{d \langle \epsilon^2 \rangle}{dt} 
&=& - \frac{2}{\tau} 
\left(1- \frac{\alpha^2}{\tau} \right) \langle \epsilon^2 \rangle
+ \left( \frac{\beta}{\tau} \right)^2, 
\hspace{1cm}\mbox{(model B)} 
\end{eqnarray}
Equations of motion for $\mu$ and $\gamma$ 
are given by Eqs. (33), (34), (A16) and (A17):
\begin{eqnarray}
\frac{d \mu}{dt} &=& f_0 + f_2 \gamma + I(t),
\hspace{1cm}\mbox{(models A and B)} \\
\frac{d \gamma}{dt} &=& 2 f_1 \gamma 
+ \frac{r_q^2 \phi^2}{(1-r_q \tau f_1)}, 
\hspace{1cm}\mbox{(model A)} \\
\frac{d \gamma}{dt} &=& 2 f_1 \gamma 
+ \frac{u^2_q \:\beta^2}{(1-\tau f_1)},
\hspace{1cm}\mbox{(model B)} 
\end{eqnarray}
where $u_q=\sqrt{(3-q)/(5-3 q)}$ [Eq. (A18)].
In the model A, we have adopted the approximation:
$K(\epsilon) \simeq -\epsilon/r_q$ [Eq. (9)],
without which reasonable results are not obtainable
in the moment approach (see the discussion below).
Equations (93)-(97) show that equations of motion 
for the models A and B have the same structure.
In the case of weak noise and small $\tau$,
for which the second-moment approach is expected to be valid,
the dynamical properties of the models A and B
(as well as the model A$_0$) are qualitatively the same,
although there are some quantitative
difference among them: 
{\it i.e.} the stationary value of $\gamma$ of the model A
is different from that of the model B.

Our discussion presented in this paper
is based on the second-order moment method.
Effects of higher-order moment neglected in our method
are examined in the following.
The equation of motion
for the $k$th moment with even $k$
of the model A$_0$ is formally given by
\begin{eqnarray}
\frac{d \langle \epsilon^k \rangle }{d t}
&=& \left( \frac{k}{\tau} \right) 
\langle \epsilon^{(k-1)} K(\epsilon) \rangle 
+ \frac{k(k-1)}{2}\left( \frac{\phi}{\tau} \right)^2
\langle \epsilon^{(k-2)} \rangle,
\hspace{1cm}\mbox{(model A$_0$)} 
\end{eqnarray}
though an evaluation of the first term of Eq. (98)
is very difficult. 
In order to get a meaningful result within the moment method,
we have assumed 
$K(\epsilon) \simeq -\epsilon/r_q$ (the model A) to get
\begin{eqnarray}
\frac{d \langle \epsilon^k \rangle }{d t}
&=& - \left( \frac{k}{r_q \tau} \right)
\langle \epsilon^{k} \rangle 
+ \frac{k(k-1)}{2}\left( \frac{\phi}{\tau} \right)^2
\langle \epsilon^{(k-2)} \rangle,
\hspace{1cm}\mbox{(model A)} 
\end{eqnarray}
from which we may recurrently calculate 
the stationary $k$th moment as
\begin{eqnarray}
\langle \epsilon^{k} \rangle 
&=& \frac{(k-1) r_q}{2} \left( \frac{\phi^2}{\tau} \right)
\langle \epsilon^{(k-2)} \rangle, \\
&=& \frac{(k-1)!! \: r_q^{k/2}}{2^{k/2}}
\left(\frac{\phi^2}{\tau}  \right)^{k/2}.
\hspace{1cm}\mbox{(model A)}  
\end{eqnarray}

The stationary distribution in the model A$_0$
given by Eq. (4)
leads to the second- and fourth-order moments:
\begin{eqnarray}
\langle \epsilon^2 \rangle 
&=& \frac{\phi^2}{(5-3 q)\tau}, \\
\langle \epsilon^4 \rangle 
&=& \frac{3 \phi^4}{ (5-3q)(7-5q)\tau^2}.
\hspace{1cm}\mbox{(model A$_0$)} 
\end{eqnarray}
In contrast, the stationary distribution in the model A
given by Eq. (92) yields
\begin{eqnarray}
\langle \epsilon^2 \rangle 
&=& \frac{(2-q)\phi^2}{(5-3 q)\tau}, \\
\langle \epsilon^4 \rangle 
&=& 3 \langle \epsilon^2 \rangle^2
= \frac{3 (2-q)^2 \phi^4}{ (5-3q)^2 \tau^2}.
\hspace{1cm}\mbox{(model A)} 
\end{eqnarray}
The expression of Eq. (104) is different
from that of Eq. (102) by a factor of $(2-q)$.
The ratio of Eq. (105) to Eq. (103) becomes
$(2-q)^2(7-5q)$, which is less than unity for $1 < q < 5/3$.
These show that the distribution
given by Eq. (92) in the model A
underestimates the effective
width of the distribution of $\epsilon$ compared to that
in the model A$_0$.
In order to include the higher-order moment in an appropriate way, 
we have to go beyond the approximation 
with $K(\epsilon) \simeq - \epsilon/r_q $ [Eq. (9)].

With the model B, we may obtain the equation of motion
for the $k$th moment with even $k$, as given by
\begin{eqnarray}
\frac{d \langle \epsilon^k \rangle }{d t}
&=& -\left[\frac{k}{\tau} 
-\frac{k^2}{2} \left( \frac{\alpha}{\tau} \right)^2 \right]
\langle \epsilon^k \rangle
+ \frac{k(k-1)}{2}\left( \frac{\beta}{\tau} \right)^2
\langle \epsilon^{(k-2)} \rangle.
\hspace{1cm}\mbox{(model B)} 
\end{eqnarray}
The stationary value of the $k$th moment is given by
\begin{eqnarray}
\langle \epsilon^k \rangle
&=& \frac{(k-1) \beta^2}{2(\tau - k \alpha^2/2)}
\langle \epsilon^{(k-2)} \rangle, \\
&=& \frac{(k-1)!!\:\beta^k}
{2^{k/2} \: \Pi_{\ell=1}^{k/2}(\tau - \ell \alpha^2)}.
\end{eqnarray}
For example, second- and fourth-moments are given by
\begin{eqnarray}
\langle \epsilon^2 \rangle 
&=& \frac{\beta^2}{2(\tau-\alpha^2)}, \\
\langle \epsilon^4 \rangle 
&=& \frac{3 \beta^4}{4(\tau-2\alpha^2)(\tau-\alpha^2)},
\hspace{1cm}\mbox{(model B)} 
\end{eqnarray}
which agree with the result obtained from
the stationary distribution given by Eq. (66) or (67).
We get the positive definite 
$\langle \epsilon^{k} \rangle$ for $\alpha^2 < 2 \tau/k$. 
This suggests that for $2 \tau/k < \alpha^2 < \tau $ 
with $k \geq 4$, the $k$th moment diverges even if 
the second moment remains finite.
This might throw some doubt on the validity of the
second-moment approach. 
Equation (106) expresses
that the motion of $\langle \epsilon^{k} \rangle$
depends on those of its lower moments ($\leq k-2$),
but it is independent of its higher moments ($\geq k+2$).
For example, even if $\langle \epsilon^{4} \rangle$
diverges, it has no effects on the motion of
$\langle \epsilon^{2} \rangle$ for $\tau/2 < \alpha^2 < \tau$.
It is promising to take into account
contributions from higher-order moments in the model B,
although its validity range becomes narrower
because $\alpha$ has to satisfy the condition:
$\alpha^2 < 2 \tau/k $ for the $k$th moment 
to remain finite.

Our discussions presented in the preceding sections
are confined to the stationary properties of the Langevin model
subjected to non-Gaussian noise. It is possible
to discuss its dynamics,
by solving equations of motion for $\mu$ and $\gamma$.
Numerical calculations for the model B are plotted in
Figure 6(a) and 6(b), which show the time dependences of
$\mu$ and $\gamma$, respectively.
We apply an external pulse
input given by $I(t)=A \:\Theta(t-100)\Theta(200-t)$
with $A=1.0$, which is plotted at the bottom
of Fig. 6(a), $\Theta(x)$ denoting the Heaviside function.
Figure 6(a) shows that $\mu(t)$ of the moment method
is in good agreement with the result of DS.
Figure 6(b) shows that $\gamma(t)$ is independent of
an input pulse [Eq. (A17)].
With increasing $\alpha$, a steady value of $\gamma$
is increased. The result of the moment method for $\alpha=0.0$
is in good agreement with that of DS although
for $\alpha=0.5$, the former is underestimated
compared to the latter.

The overall behavior of the stationary distribution  
is fairly well reproduced by all the approximations mentioned in Sec. 2.2. 
Tails of $P(x)$ are, however, not satisfactorily described, 
in particular, in calculations of the model A.
This is partly due to the fact that the approximate
Eq. (10) yields the Gaussian stationary distribution 
given by Eq. (92),
though Eqs. (2) and (3) are originally introduced to
generate non-Gaussian noise.
This point is improved in the model B, in which
the stationary distribution given by Eq. (64) is non-Gaussian
as expressed by Eqs. (66) and (67).
Indeed, tails of $P(x)$ of Fig. 5 for the model B
are slightly well reproduced than those of Fig. 3 for the model A.

\section{Conclusion}
To summarize,
we have studied effects of non-Gaussian noise on the
Langevin model, by using the second-order moment approach.
The obtained result is summarized as follows.

\noindent
(1) With increasing $\tau$, the width of the 
stationary distribution $P(x)$ is decreased.

\noindent
(2) For $q>1$ ($q < 1$), the width of $P(x)$ is 
increased (decreased) compared to that for $q=1$.

\noindent
(3) The prefactor of $F_{eff}$ for the model A
in the moment method agrees with that in FI-1, 
but disagrees with that in the UCNA and FI-2 (Table 1).

\noindent
The items (1) and (2) are realized in both the
models A and B.  This may be explained by the $q$-
and $\tau$-dependent $\alpha_{eff}$ [Eq. (38)]
or $\beta_{eff}$ [Eq. (78)]. 
As for the item (3), it is necessary to point out that
although the UCNA \cite{Jung87,Hanggi95} exactly interpolates
between the two limits of $\tau=0$ and $\tau=\infty$, 
it is not exact for $O(\tau)$ \cite{Mallick06}.
The functional integral method is a formally exact transformation
if the functional integral is correctly performed. 
In the actual applications, however, it is inevitable to adopt
some kinds of approximation, with which the final result
depends on the adopted approximation.
The difference between the results of 
FI-1 \cite{Wu07} and FI-2 \cite{Fuentes01} arise
from the difference between the adopted approximations
in performing the functional integral.
These yield the difference in the results
listed in the Table 1.

As for the models A and B, we get

\noindent
(i) although the stationary distribution of $p(\epsilon)$
in the model A is the Gaussian, 
the effect of the non-Gaussian distribution
of the original model A$_0$ is fairly well taken into
account by a factor of $r_q$, and

\noindent
(ii) the newly introduced model B, which yields
the stationary non-Gaussian
$p(\epsilon)$ equivalent to that of
the model A$_0$, is expected to be a promising model
generating non-Gaussian noise.

It is possible to apply the moment approach to
a wide class of stochastic systems subjected to non-Gaussian noise,
because its calculation is simple and transparent.
It would be interesting to investigate effects of non-Gaussian noise
on the synchronization in coupled nonlinear
systems with the use of the model B, 
which is left as our future study.

\section*{Acknowledgements}
This work is partly supported by
a Grant-in-Aid for Scientific Research from the Japanese 
Ministry of Education, Culture, Sports, Science and Technology.  

\vspace{1cm}
\newpage

\appendix

\noindent
{\bf\large Appendix}

We discuss an application of the moment method
to the model B given by Eqs. (63) and (64), for which
the FPE of the distribution $p(x, \epsilon, t)$
in the Stratonovich representation is given by
\begin{eqnarray}
\frac{\partial }{\partial t}p(x,\epsilon,t) &=& 
-\frac{\partial }{\partial x}
\{[F(x)+ \epsilon + I] p(x,\epsilon,t)\} 
+\frac{1}{\tau}\frac{\partial}{\partial \epsilon} 
[\epsilon p(x,\epsilon,t)] 
\nonumber \\
&+& \frac{1}{2}\left(\frac{\alpha}{\tau}\right)^2 
\frac{\partial}{\partial \epsilon} \left[\epsilon
\frac{\partial}{\partial \epsilon} \epsilon p(x,\epsilon,t)\right]
+ \frac{1}{2}\left(\frac{\beta}{\tau}\right)^2 
\frac{\partial^2}{\partial \epsilon^2}
p(x,\epsilon,t).
\nonumber \hspace{1cm}\mbox{(A1)}
\end{eqnarray}
We define means, variances and covariances by
\begin{eqnarray}
\langle x^m \epsilon^n \rangle 
&=& \int dx \int d \epsilon \; x^m \epsilon^n p(x,\epsilon,t).
\hspace{1cm}\mbox{($m, n$: integer)}
\nonumber \hspace{2.5cm}\mbox{(A2)}
\end{eqnarray}
After simple calculations using Eqs. (A1) and (A2),
we get their equations of motion given by \cite{Hasegawa06,Hasegawa07}
\begin{eqnarray}
\frac{d \langle x \rangle}{d t} 
&=& \langle F(x) + \epsilon + I \rangle,
\nonumber \hspace{7cm}\mbox{(A3)} \\
\frac{d \langle \epsilon \rangle}{d t} 
&=& - \frac{1}{\tau}\langle  \epsilon \rangle
+\frac{1}{2}\left(\frac{\alpha}{\tau}\right)^2\langle \epsilon \rangle, 
\nonumber \hspace{6cm}\mbox{(A4)} \\
\frac{d \langle x^2 \rangle}{d t} 
&=& 2 \langle x[F(x) + \epsilon + I] \rangle, 
\nonumber \hspace{6.5cm}\mbox{(A5)} \\
\frac{d \langle \epsilon^2 \rangle}{d t} 
&=& - \frac{2}{\tau}\langle \epsilon^2 \rangle
+ 2 \left(\frac{\alpha}{\tau} \right)^2 \langle \epsilon^2 \rangle 
+\left(\frac{\beta}{\tau} \right)^2,
\nonumber \hspace{4cm}\mbox{(A6)} \\
\frac{d \langle x \epsilon \rangle}{dt}
&=& \langle \epsilon [F(x) + \epsilon + I] \rangle
-\frac{1}{\tau}\langle x \epsilon \rangle.
\nonumber \hspace{5cm}\mbox{(A7)}
\end{eqnarray}
We will consider the variables of
$\mu$, $\nu$, $\gamma$, $\zeta$ and $\phi$
defined by Eqs. (19)-(23).
Their equations of motion become  \cite{Hasegawa06,Hasegawa07}
\begin{eqnarray}
\frac{d \mu}{dt} &=& f_0 + f_2 \gamma 
+ \nu  +I(t), 
\nonumber \hspace{6cm}\mbox{(A8)} \\
\frac{d \nu}{dt} &=& - \frac{\nu}{\tau} 
+\frac{1}{2}\left(\frac{\alpha}{\tau}\right)^2 \nu,
\nonumber \hspace{7cm}\mbox{(A9)} \\
\frac{d \gamma}{dt} &=& 2(f_1 \gamma + \phi), 
\nonumber \hspace{8cm}\mbox{(A10)} \\
\frac{d \zeta}{dt} &=& - \frac{2}{\tau} \zeta
+ 2 \left(\frac{\alpha}{\tau}\right)^2 \zeta
+ \nu^2 \left(\frac{\alpha}{\tau} \right)^2 
+ \left(\frac{\beta}{\tau} \right)^2,
\nonumber \hspace{3cm}\mbox{(A11)} \\
\frac{d \phi}{dt} &=& \left(-\frac{1}{\tau} + f_1 \right) \phi
+ \zeta-\frac{\mu \nu}{2}\left(\frac{\alpha}{\tau} \right)^2,
\nonumber \hspace{4.5cm}\mbox{(A12)}
\end{eqnarray}
where $f_{\ell}=(1/\ell !) \partial^{\ell} F(\mu)/\partial x^{\ell}$.

When we assume the stationary values
for $\nu$, $\zeta$ and $\phi$:
\begin{eqnarray}
\nu &\simeq & \nu_s = 0,
\nonumber \hspace{8.5cm}\mbox{(A13)} \\
\zeta &\simeq & \zeta_s = \frac{\beta^2}{2(\tau-\alpha^2)},
\nonumber \hspace{7cm}\mbox{(A14)} \\
\phi &\simeq& \phi_s = \frac{\beta^2}{2(1-\alpha^2/\tau)(1-\tau f_1)},
\nonumber \hspace{5cm}\mbox{(A15)}
\end{eqnarray}
equations of motion for 
$\mu$ and $\gamma$ become
\begin{eqnarray}
\frac{d \mu}{dt} &=& f_0 + f_2 \gamma + I(t),
\nonumber \hspace{7cm}\mbox{(A16)} \\
\frac{d \gamma}{dt} &=& 2 f_1 \gamma 
+ \frac{u^2_q \:\beta^2}{(1-\tau f_1)},
\nonumber \hspace{6.5cm}\mbox{(A17)} 
\end{eqnarray}
with
\begin{eqnarray}
u_q &=& \sqrt{\frac{1}{(1-\alpha^2/\tau)}}
= \sqrt{\frac{3-q}{5-3q}}.
\nonumber \hspace{5cm}\mbox{(A18)}
\end{eqnarray} 
With increasing $\tau$, $u_q$ is decreased 
because of a decreased $q$ (Fig. 1).
Equations (A17) and (A18) lead to the stationary value of $\gamma_s$
given by
\begin{eqnarray}
\gamma_s
&=& \frac{\beta^2(3-q) }{(-2 f_1)(1-\tau f_1)(5-3q)}
= \frac{\tau}{(-f_1)(1-\tau f_1)} \langle \epsilon^2 \rangle.
\nonumber \hspace{2cm}\mbox{(A19)}
\end{eqnarray}
It is noted that equations of motion given by
Eqs. (A16) and (A17) may be derived from
the one-variable DE given by Eq. (76) \cite{Hasegawa06,Hasegawa07}.

\newpage

\begin{center}
\begin{tabular}{|c|c|c|c|}
\hline
$F_{eff}$ & $I_{eff}$ & $\alpha_{eff}$ & method \\ \hline \hline
$F$ & $I$ & $r_q \phi/(\sqrt{1-r_q \tau f_1})$ 
& ${\rm moment}^{1)}$ \\
$F$ & $-$ & $r_q \phi/(\sqrt{1-r_q \tau F'_s})$ 
& FI-1$^{2)}$ \\
$F/(1-r_q \tau F')$ & $(I+\tau \dot{I})/(1-r_q \tau F')$ 
& $r_q \phi/(1-r_q \tau F')$ & ${\rm UCNA}^{3)}$ \\
$F/(1- s_q \tau F')$ & $-$ & $s_q \phi /(1- s_q \tau F') $
& FI-2$^{4)}$ \\

\hline
\end{tabular}
\end{center}

{\it Table 1} A comparison among various approaches
to the model A [Eqs. (1) and (10)]
yielding the effective differential equation given by 
$\dot{x}= F_{eff}+ I_{eff} + \alpha_{eff}\: \eta(t)$, where
$r_q=2(2-q)/(5-3q)$ and $s_q=1+(q-1)(\tau/2\phi^2)F^2$;
(1) the moment method: 
(2) functional-integral (FI-1) method of Ref. \cite{Wu07}: 
(3) UCNA calculation after Ref. \cite{Jung87,Hanggi95}:
(4) functional-integral (FI-2) method of Ref. \cite{Fuentes01} (see text).

\newpage



\newpage

\begin{figure}
\caption{
The $q$ dependence of $r_q$ [Eq. (11): solid curve]
and $u_q$ [Eq. (80): dashed curve].
}
\label{fig1}
\end{figure}

\begin{figure}
\caption{
The stationary probability $P(x)$ 
for $F(x)=-\lambda x$ of the model A$_0$ [Eqs. (1)-(3)] 
calculated by direct simulation (DS: dashed curves),
and $P(x)$ of the model A [Eqs. (1) and (10)] calculated by
the moment method (solid curves),
UCNA (chain curves) and FI-2 (dotted cures):
results of FI-1 agree with those of the moment method:
(a) $(q, \tau)=( 0.8, 0.5)$, 
(b) $(0.8, 1.0)$, 
(c) $(1.0, 0.5)$, 
(d) $(1.0, 1.0)$, 
(e) $(1.5, 0.5)$, and
(f) $(1.5, 1.0)$
($\phi=0.5$ and $\psi=0.0$).
}
\label{fig2}
\end{figure}

\begin{figure}
\caption{
The stationary probability $P(x)$ 
for $F(x)= x - x^3$ of the model A$_0$ [Eqs. (1)-(3)] 
calculated by direct simulation (DS: dashed curves),
and $P(x)$ of the model A [Eqs. (1) and (10)] calculated by
the moment method (solid curves),
UCNA (chain curves) and FI-2 (dotted cures):
results of FI-1 agree with those of the moment method:
(a) $(q, \tau)=( 0.8, 0.5)$, 
(b) $(0.8, 1.0)$, 
(c) $(1.0, 0.5)$, 
(d) $(1.0, 1.0)$, 
(e) $(1.5, 0.5)$, and
(f) $(1.5, 1.0)$
($\phi=0.5$ and $\psi=0.0$).
}
\label{fig3}
\end{figure}

\begin{figure}
\caption{
The stationary probability $P(x)$ 
for $F(x)=-\lambda x$ of the model B given by Eqs. (63) and (64)
with
(a) $(\alpha, \tau)=( 0.0, 0.5)$ ($q=1.40$), 
(b) $(0.0, 1.0)$ ($q=1.22$), 
(c) $(0.5, 0.5)$ ($q=1.40$), and
(d) $(0.5, 1.0)$ ($q=1.22$), 
calculated for $\beta=0.5$
by DS (dashed curves) and
the moment method (solid curves).
}
\label{fig4}
\end{figure}

\begin{figure}
\caption{
The stationary probability $P(x)$ 
for $F(x)=x-x^3$ of the model B given by Eqs. (63) and (64)
with
(a) $(\alpha, \tau)=( 0.0, 0.5)$ ($q=1.40$), 
(b) $(0.0, 1.0)$ ($q=1.22$), 
(c) $(0.5, 0.5)$ ($q=1.40$), and
(d) $(0.5, 1.0)$ ($q=1.22$), 
calculated for $\beta=0.5$ by DS (dashed curves) and
the moment method (solid curves).
}
\label{fig5}
\end{figure}

\begin{figure}
\caption{
The time dependence of
(a) $\mu(t)$ and (b) $\gamma(t)$ of the model B
for $\alpha=0.0$ and $\alpha=0.5$
with $\beta=0.5$ and $\tau=1.0$,
calculated by the moment method (solid curves)
and DS (dashed curves),
an input pulse being plotted at the bottom of (a).  
Results for $\alpha=0.5$ in (a) is indistinguishable
from those for $\alpha=0.0$.
}
\label{fig6}
\end{figure}

\end{document}